\documentclass[oldversion]{aa}  
\usepackage{thumbpdf}
\usepackage{amsmath}
\usepackage{graphicx}
\usepackage{natbib}
\bibpunct{(}{)}{;}{a}{}{,}
\usepackage{txfonts}
\usepackage{longtable}
\usepackage{supertabular}
\begin{document}
\bibliographystyle{apj}
%
%  These Macros are taken from the AAS TeX macro package version 4.0.
%  Include this file in your LaTeX source only if you are not using
%  the AAS TeX macro package and need to resolve the macro definitions
%  in the BibTeX entries returned by the ADS abstract service.
%
%  For more information on the AASTeX macro package, please see the URL
%	http://www.aas.org/publications/aastex.html
%  For more information about ADS abstract server, please see the URL
%	http://adswww.harvard.edu/ads_abstracts.html
%

% Abbreviations for journals.  The object here is to provide authors
% with convenient shorthands for the most "popular" (often-cited)
% journals; the author can use these markup tags without being concerned
% about the exact form of the journal abbreviation, or its formatting.
% It is up to the keeper of the macros to make sure the macros expand
% to the proper text.  If macro package writers agree to all use the
% same TeX command name, authors only have to remember one thing, and
% the style file will take care of editorial preferences.  This also
% applies when a single journal decides to revamp its abbreviating
% scheme, as happened with the ApJ (Abt 1991).

\def\jnl@style{\it}
%commente par Seb
\def\aaref@jnl#1{{\jnl@style#1}}
%ref remplace par aaref pour eviter conflit...

\def\aaref@jnl#1{{\jnl@style#1}}

\def\aj{\aaref@jnl{AJ}}                   % Astronomical Journal
\def\araa{\aaref@jnl{ARA\&A}}             % Annual Review of Astron and Astrophys
\def\apj{\aaref@jnl{ApJ}}                 % Astrophysical Journal
\def\apjl{\aaref@jnl{ApJ}}                % Astrophysical Journal, Letters
\def\apjs{\aaref@jnl{ApJS}}               % Astrophysical Journal, Supplement
\def\ao{\aaref@jnl{Appl.~Opt.}}           % Applied Optics
\def\apss{\aaref@jnl{Ap\&SS}}             % Astrophysics and Space Science
\def\aap{\aaref@jnl{A\&A}}                % Astronomy and Astrophysics
\def\aapr{\aaref@jnl{A\&A~Rev.}}          % Astronomy and Astrophysics Reviews
\def\aaps{\aaref@jnl{A\&AS}}              % Astronomy and Astrophysics, Supplement
\def\azh{\aaref@jnl{AZh}}                 % Astronomicheskii Zhurnal
\def\baas{\aaref@jnl{BAAS}}               % Bulletin of the AAS
\def\jrasc{\aaref@jnl{JRASC}}             % Journal of the RAS of Canada
\def\memras{\aaref@jnl{MmRAS}}            % Memoirs of the RAS
\def\mnras{\aaref@jnl{MNRAS}}             % Monthly Notices of the RAS
\def\pra{\aaref@jnl{Phys.~Rev.~A}}        % Physical Review A: General Physics
\def\prb{\aaref@jnl{Phys.~Rev.~B}}        % Physical Review B: Solid State
\def\prc{\aaref@jnl{Phys.~Rev.~C}}        % Physical Review C
\def\prd{\aaref@jnl{Phys.~Rev.~D}}        % Physical Review D
\def\pre{\aaref@jnl{Phys.~Rev.~E}}        % Physical Review E
\def\prl{\aaref@jnl{Phys.~Rev.~Lett.}}    % Physical Review Letters
\def\pasp{\aaref@jnl{PASP}}               % Publications of the ASP
\def\pasj{\aaref@jnl{PASJ}}               % Publications of the ASJ
\def\qjras{\aaref@jnl{QJRAS}}             % Quarterly Journal of the RAS
\def\skytel{\aaref@jnl{S\&T}}             % Sky and Telescope
\def\solphys{\aaref@jnl{Sol.~Phys.}}      % Solar Physics
\def\sovast{\aaref@jnl{Soviet~Ast.}}      % Soviet Astronomy
\def\ssr{\aaref@jnl{Space~Sci.~Rev.}}     % Space Science Reviews
\def\zap{\aaref@jnl{ZAp}}                 % Zeitschrift fuer Astrophysik
\def\nat{\aaref@jnl{Nature}}              % Nature
\def\iaucirc{\aaref@jnl{IAU~Circ.}}       % IAU Cirulars
\def\aplett{\aaref@jnl{Astrophys.~Lett.}} % Astrophysics Letters
\def\apspr{\aaref@jnl{Astrophys.~Space~Phys.~Res.}}
                % Astrophysics Space Physics Research
\def\bain{\aaref@jnl{Bull.~Astron.~Inst.~Netherlands}} 
                % Bulletin Astronomical Institute of the Netherlands
\def\fcp{\aaref@jnl{Fund.~Cosmic~Phys.}}  % Fundamental Cosmic Physics
\def\gca{\aaref@jnl{Geochim.~Cosmochim.~Acta}}   % Geochimica Cosmochimica Acta
\def\grl{\aaref@jnl{Geophys.~Res.~Lett.}} % Geophysics Research Letters
\def\jcp{\aaref@jnl{J.~Chem.~Phys.}}      % Journal of Chemical Physics
\def\jgr{\aaref@jnl{J.~Geophys.~Res.}}    % Journal of Geophysics Research
\def\jqsrt{\aaref@jnl{J.~Quant.~Spec.~Radiat.~Transf.}}
                % Journal of Quantitiative Spectroscopy and Radiative Transfer
\def\memsai{\aaref@jnl{Mem.~Soc.~Astron.~Italiana}}
                % Mem. Societa Astronomica Italiana
\def\nphysa{\aaref@jnl{Nucl.~Phys.~A}}   % Nuclear Physics A
\def\physrep{\aaref@jnl{Phys.~Rep.}}   % Physics Reports
\def\physscr{\aaref@jnl{Phys.~Scr}}   % Physica Scripta
\def\planss{\aaref@jnl{Planet.~Space~Sci.}}   % Planetary Space Science
\def\procspie{\aaref@jnl{Proc.~SPIE}}   % Proceedings of the SPIE

\let\astap=\aap
\let\apjlett=\apjl
\let\apjsupp=\apjs
\let\applopt=\ao

   \title{Dust content of core-collapse supernova hosts}

%   \subtitle{}

   \author{A.-L. Melchior
          \inst{1,2}
          \and
          F. Combes\inst{1}
          }

   \offprints{}

   \institute{LERMA, Observatoire de Paris, CNRS, UMR8112, 61, avenue de
l'Observatoire, F-75014 Paris, France\\ 
              \email{A.L.Melchior@obspm.fr, Francoise.Combes@obspm.fr}
         \and
	   Universit\'e Pierre et Marie Curie-Paris 6, 4, Place Jussieu,
F-75\,252 Paris Cedex 05, France
             }

   \date{}

   \abstract{We study a small sample of $z=0.1-0.6$ core-collapse
supernova (CCSN) host galaxies. {Continuum observations at 250GHz have
been performed with MAMBO at the IRAM-30m telescope.  None
of these sources has been detected and the error-weighted mean flux is
0.25$\pm$0.32\,mJy.  Upper limits on their dust masses are derived and
the corresponding sample mean corresponds to $1.4\pm 2.2 \times 10^8
M_\odot$. These results are comparable with previous submillimetre
observations of SN-Ia hosts performed by Farrah et al. and by Clements
et al.  We conclude that CCSN hosts are not extreme at
millimetre wavelengths, and as confirmed with the optical
luminosities of a subset of our sample, they are typical of the local
galaxy population. }
} \maketitle
%
%________________________________________________________________
\section{Introduction} 
\label{sec:intro}
In the Local Universe, type-Ia supernovae (SN-Ia) are detected in host
galaxies along the whole Hubble sequence, while the core-collapse
supernovae (CCSN) tend to avoid early-type galaxies (without
gas). Star formation activity is known to increase with redshift
\citep[e.g.][]{Madau:1996,Madau:1998a,Chary:2001,Blain:2002}, while no
evolution is observed in supernova properties as yet
\citep[e.g.][]{Combes:2004}.  Submillimetre and millimetre (submm/mm) galaxies
(SMG) account for a significant fraction of the submm/mm Cosmic
Background
\citep[e.g.][]{Hughes:1998,Barger:1999}, and 50-70$\%$
have {a radio} counterpart {(of the order of 100$\mu$Jy). Most
of these SMG are detected at $z\sim 2-3$}
\citep[e.g.][]{Smail:2000,Chapman:2003a,Chapman:2003b,Chapman:2005}.
{The submm/mm fluxes thus correspond to the redshifted far-infrared
(FIR) peak of the dust emission.}  Our current understanding of {the
FIR/radio} correlation {\citep{Condon:1992}} is that synchrotron
emission detected in radio is due to relativistic electrons
accelerated by shocks during supernova explosions. {Numerous works
support this interpretation of the FIR/radio emission in term of a
recent star formation activity
\citep[e.g.][]{Boyle:2007,Vlahakis:2007}, pointing towards a strong
link between SMG and strong starbursts.}  In addition,
\cite{Dunne:2003} have shown that CCSN are at least as important as
stellar winds enriching our Galaxy with dust and suggested that they
could be the dominant source of dust at high redshift. The role of
supernovae and CCSN should thus be essential in these
mechanisms. However, various difficulties still prevent definite
conclusions. (1) {In the optical,} extinction biases the detection
of CCSN in the core of strong starbursts \citep{Mattila:2001}:
\citet{Mattila:2007} have detected for the first time a CCSN within a
circumnuclear starburst suffering 40\,mag. of extinction in V. (2)
There is still no direct evidence that CCSN could be dust factories at
high-z \citep{Meikle:2007}.  {The rates of radio supernovae
\citep[e.g.][]{Lonsdale:2006} and supernova remnants
\citep[e.g.][]{Lenc:2006} do not address the first question, but only
probe nearby starbursts so far.}
\begin{table*}
\caption{List of the CCSN galaxies observed for 250GHz-MAMBO continuum at
IRAM-30m. 
(1): IAU designation; (2,3): J2000 RA, DEC coordinates; (4):
redshift of the host galaxy or the SN; (5): SN type; (6): detection
magnitude of the SN; (7): IAU circular numbers.
}
\centering
\begin{tabular}{llllllll}
\hline\hline
Name & RA & DEC & z & type & Mag. & IAUC \\ 
(1) & (2) & (3) & (4) & (5) & (6) & (7) \\
SN & J2000 & J2000 & & & & \\\hline
1995av& 02 01 36.7 & +03 38 55.2 & 0.30 &    II& 20.1 & 6270\\
1999fl & 02 30 05.5 & +00 44 52.6 & 0.3   &    II & 24.9 & 7312\\
2001ek$^{a}$ &02 30 17.6 & +01 03 56.4& 0.25-0.40 & IIP & 25.1 & 7719\\
1999fp& 04 15 02.5&  +04 21 46.4&  0.34  &      II? & 24.1 & 7312 \\
2000ei  &04 17 07.2 &+05 45 53.1& 0.60    &   II?  & 22.8 & 7516\\
1997ev& 08 24 20.2 & +03 51 36.0&  0.43  &    II?& 23.0 & 6804\\
1997ew &08 24 25.0&  +03 49 08.0 &  0.59 &   II/Ic?& 23.9 &6804 \\
1998at & 10 54 54.3 & $-$03 44 10.0 & 0.40 & II  & 23.8 &  6881\\
2001ct &13 24 44.6 & +27 15 27.2 & 0.45  & II?& 23.4 & 7649\\
2002cm &13 52 03.7&  $-$11 43 08.5 & 0.087   &  I& 22.3 & 7885\\ 
2002du &13 53 18.3&  $-$11 37 28.8&  0.21 & II & 22.2 & 7929\\
2001gl &14 01 16.6&  +05 12 48.9& 0.36  & Ib/c & 23.7 & 7763\\
2002co &14 10 53.0 & $-$11 45 25.0 & 0.318  &   II & 22.6 &7885\\
\hline
\end{tabular}
 \\$^{a}$ We assume
$z=0.25$ in Figure \ref{fig:snmm}.
\label{tab:sources}
\end{table*}

{In the local Universe, the strongest starbursts are found in
interacting and merging galaxies \citep{Sanders:1996}, while
\citet{Conselice:2003} suggested that two thirds of the SMG at $z>1$
are undergoing a major merger. These observations are in good
agreement with the hierarchical scenario, even though all mergers do
not trigger a strong starburst
\citep[e.g.][]{Bergvall:2003,DiMatteo:2007}.  While the dust attenuation is
expected to increase with the bolometric luminosity of the galaxies
and their SFR, the evolution of the dust attenuation with redshift is
still a matter of debate \citep[e.g.][]{Buat:2007}.}  Nevertheless,
SMG observations suggest that these young galaxies contain a large
fraction of gas {\citep[e.g.][]{Neri:2003,Greve:2005,Tacconi:2006},
 probably} associated with a large amount of dust.

{The variation with redshift of the mean extinction affecting hosts}
is {a real} concern for the high-z SN searches. {\citet{Farrah:2004}
tentatively argue that the extinction in high-z SN-Ia hosts may
increase with redshift
\citep[see also e.g.][]{Totani:1999}.} Selection effects are usually
thought to {favour hosts with less extinction}
\citep[e.g.][]{Commins:2004}, as very extinguished supernovae are not
detected in the actual surveys. {In addition, several SN-Ia
surveys apply cuts on the ($A_V \le 0.5-1$) extinction \citep[see
e.g.][]{Tonry:2003,Riess:2004,Riess:2005,Riess:2007}, while moderate
extinction is removed implicitly in the MLCS and $\Delta m_{15}$
reduction methods \citep{Riess:1996,Hamuy:1996}. There are indications
that the distant SN-Ia might have a somewhat bluer colour than their
local counterparts \citep[][and references
therein]{Leibundgut:2001}. Last, } \citet{Chary:2005} have found with
Spitzer MIPS observations of the GOODS fields that supernova host
galaxies have a detection rate at 24\,$\mu$m that is a factor of 1.5
higher than the field galaxy population. Their sample is based on 50
supernovae sampled up to $z\sim\,1$ and surprisingly, they find
similar properties for SN-Ia and CCSN hosts.

Farrah et al. (2004) have statistically detected at the $3\sigma$ level
the continuum emission at 350\,GHz of a sample of 16 SN-Ia hosts at
$z=0.5$, while one submm strong source has been detected directly. The
authors interpret this result as the signature of a slight increase
(25$\%$-135$\%$) of the dust content of this galaxy sample with
respect to local {normal} galaxies, {studied by
\citet{Rowan-Robinson:2003}.} {However, one should note that the
definition of a ``normal galaxy'' reference sample is one difficulty
of this type of work, which imposes magnitude thresholds and requires
infrared/ultraviolet detections.}

In this paper, we perform a study of the 1.2\,mm continuum on a sample
of 13 CCSN.  In Sect. \ref{sect:obs}, the observations performed at
the IRAM-30m telescope are presented together with the
data reduction with the MOPSI software. In Sect. \ref{sect:resu}, we
discuss our negative results.

Throughout this paper, we adopt a flat cosmology with
{$\Omega_{m}=0.3$, $\Omega_\Lambda=0.7$ and
$H_0=70$~km\,s$^{-1}$\,Mpc$^{-1}$.}
\begin{table*}
\caption{List of the CCSN galaxies observed for 250GHz-MAMBO continuum at
IRAM-30m.(1): IAU designation; (2): zenith optical depth of the
atmosphere at 250GHz; (3): measured flux and 1$\sigma$ error of the
sources; (4): dust mass assuming T$_d=15$K; (5): dust mass assuming T$_d=20$K; (6): date of the observation; (7): integration
time in minutes; (8): bolometer used: MAMBO-1 (37-channels) or
MAMBO-2 (117-{channels}). }
\centering
\begin{tabular}{cccccccc}
\hline\hline
Name &  $\tau_{250GHz}$ & $S(\nu_{250GHz})$ & $
{M_{dust}^{15K}}$ &$ {M_{dust}^{20K}}$ & date & IT & Bolo\\   
(1) & (2) & (3) & (4) & (5) & (6) & (7) & (8) \\ 
SN &  & mJy & ${10^9 M_\odot}$ & ${10^9 M_\odot}$ & & min & \\\hline
1995av       & 0.54-0.66 &  0.96$\pm$1.25 &{0.39$\pm$ 0.60} & {0.25$\pm$ 0.39}& 19/10/02       & 36 & 37\\
1999fl       & 0.58-0.62 & -2.63$\pm$1.89 &{-2.86$\pm$ 1.61} & {-1.77$\pm$ 1.00}& 19/10/02       & 18 & 37 \\
2001ek$^{a}$ & 0.50-0.54 & -0.22$\pm$1.48 &{ -0.18$\pm$ 0.89} & {-0.12$\pm$ 0.56}& 15/03/03       & 26 & 117 \\
1999fp       & 0.39-0.89 &  0.61$\pm$0.53 &{  0.05$\pm$ 0.10} & { 0.03$\pm$ 0.07}& 05,11/10/02    &132 & 37\\
2000ei       & 0.50-0.52 & -2.02$\pm$1.48 &{ -0.64$\pm$ 0.41} & {-0.42$\pm$ 0.27}& 01/03/03       & 25 & 117 \\
1997ev       & 0.33-0.72 &  0.33$\pm$0.59 &{  0.21$\pm$ 0.83} & { 0.14$\pm$ 0.53}& 05,11/10/02    & 93 & 37\\
1997ew       & 0.27-0.39 &  1.64$\pm$1.05 &{  0.91$\pm$ 0.81} & { 0.59$\pm$ 0.52}& 31/10,01/11/02 & 53 & 117\\
1998at       & 0.30-0.35 & -0.98$\pm$1.10 &{ -0.51$\pm$ 0.97} & {-0.33$\pm$ 0.63}& 01/11/02       & 48 & 117\\
2001ct       & 0.29-0.43 &  0.22$\pm$1.07 &{  0.13$\pm$ 0.32} & { 0.08$\pm$ 0.21}& 10-11/02/03    & 43 & 117 \\
2002cm       & 0.45-0.53 &  1.54$\pm$1.31 &{  1.12$\pm$ 0.80} & { 0.72$\pm$ 0.51}& 02,10/02/03    & 32 & 117\\ 
2002du       & 0.48-0.57 & -0.67$\pm$1.30 &{ -0.53$\pm$ 0.47} & {-0.34$\pm$ 0.30}& 02,10/02/03    & 39 & 117\\
2001gl       & 0.37-0.55 & -0.02$\pm$0.84 &{ -0.02$\pm$ 1.12} & {-0.01$\pm$ 0.70}& \multicolumn{1}{|c}{02,11/02/2003} &32 & 117\\ 
&&&&&   \multicolumn{1}{|c}{13/03/03}& 26 & 117\\	       
&&&&&   \multicolumn{1}{|c}{01/11/02}& 14 & 117\\
2002co       & 0.53-0.55&   0.77$\pm$1.46 & {0.40$\pm$ 0.64}& {0.26$\pm$ 0.42} & 02,11/02/03    & 32 & 117 \\
\hline
\end{tabular}
\\ $^{a}$ We assume $z=0.25$.
\label{tab:sourcesobs}
\end{table*}

\section{Observations and data reduction}
\label{sect:obs}
We have selected from the IAU list CCSN host galaxies with a redshift
identification available in the range $0.1<z<1$ and a secure
core-collapse type. The IAU circulars corresponding to each selected
CCSN are provided in Table
\ref{tab:sources}. 

Millimetre continuum measurements were made during the Summer-2002 and
Winter-2003 Pool observations on the MAMBO-1 (37-{channels}) and
MAMBO-2 (117-{channels}) arrays installed on the IRAM-30m telescope
(IRAM\footnote{IRAM is a joint organisation founded by the German
Max-Planck-Society, the French CNRS and the Spanish National
Geographical Institute.}, Pico Veleta, Spain). The beam size (HPBW)
was 11''. Each of the CCSN host galaxies listed in Table
\ref{tab:sources} had been observed at 250 GHz using the on-off
mode {with the most sensitive channel (resp. no 1 and 20)}. To
account for variations of the sky brightness, we used standard
chopping of the secondary mirror of the telescope between the
on-source position and a position $\pm$33'' away in azimuth, at a rate
of 2Hz. In addition, the telescope nodded {every 10 seconds such
that the previous "off" position becomes the "on" position, in order
to subtract background asymmetries between the two wobbler positions.}
Measurements of the sky opacity (sky dips) were taken approximately
every two hours, while calibration sources were regularly
monitored. The pointing was checked once per 30-60 minutes depending
on the stability {of the} observing conditions. The focus was also
regularly checked every 1-2 hours.

The data were reduced with the MOPSI software \citep{Zylka:1998}
following the standard procedure. A visual inspection of each scan had
first been performed to remove the very noisy {subscans. The
skynoise reduction used the measurements from the six channels
surrounding the photometric channel. Extinction} correction was
performed relying on a linear interpolation between sky dips taken
before and after each set of observations or the closest value.

\section{Results and discussion}
\label{sect:resu}
\subsection{Detection rate}
As presented in Table \ref{tab:sourcesobs}, we did not detect any of the
13 CCSN host galaxies we observed. We also compute the error-weighted
mean flux of the whole sample and find 0.25$\pm$0.32\,mJy. We reach a
sensitivity comparable to the sample (${1.55\pm 0.31}$\,mJy) of
\citet{Farrah:2004}. However, contrary to these authors, we do not
find any trace of signal. {Their} detection is dominated by 2
sources (SN1997ey and SN2000eh) and hence very sensitive to
statistical fluctuations. {In addition, we have shown in
\citet{Melchior:2007} that the strongest of these 2 sources (SN1997ey
host) has no significant gas content given its submillimetre flux: the
latter might be due to a background source. Excluding this source,
\citet{Farrah:2004} find an error-weighted mean flux of their 16
galaxies of $1.01\pm 0.33$\, mJy. Removing the second source,
detected at 3$\sigma$ and at 850$\mu$m only, their mean flux reduces
to 0.84$\pm$0.40\,mJy.} \citet{Clements:2005} have subsequently
extended the SN-Ia host sample to 31 galaxies and found a less
significant signal (0.44$\pm${0.22}\,mJy). Hence, we can conclude that
our CCSN host measurements are consistent with the SN-Ia host
findings. However, one can note that for a given level of rms
sensitivity, the SCUBA measurements probe much deeper the thermal
emission of galaxies as the 850-to-1200\,$\mu$m flux ratio is expected
around 3.

\subsection{Dust masses and comparison}
Deriving dust masses from mm/submm continuum measurements depends on
the complex parameters characterising the dust. They have first been
provided by Hildebrand (1983)\nocite{Hildebrand:1983}, who relied on
reflection nebulae data. Large scatters thus affect the grain
emissivity $Q(\lambda)$, the grain radius $a$ and the material density
$\rho$, as further studied by various subsequent workers
\citep[e.g.][]{Draine:1984,Casey:1991}. In addition, initial works
were based on 100/250$\mu$m measurements, which probe the hot dust
component and tend to underestimate the total dust mass. James et
al. (2002)\nocite{James:2002} has proposed a new approach to determine
the dust mass-absorption coefficient $\kappa_d (850\mu m) = 3 Q(850\mu
m)/(4a\rho)$, assuming that the fraction of metals within the
interstellar medium of a galaxy bound up in dust is constant. They
provide arguments in favour of this assumption and estimate $\kappa_d
(850\mu m)=0.07\pm0.02$m$^2$ kg$^{-1}$. This coefficient can be
extrapolated to {lower (or higher)} frequencies thanks to the
emissivity index $\beta$ usually taken in the range $\beta\sim 1-2$ :
$\kappa_d(\lambda)\propto \nu^\beta$. Following e.g. Dunne et
al. (2001)\nocite{Dunne:2001}, we considered $\beta=2$, assuming
optically thin dust emission.  The dust mass $M_{dust}$ can thus be
determined by the formula :
\begin{equation}
M_{dust} = \frac{S(\nu_{obs}) {D_L}^2}{(1+z) \kappa_d(\nu_e)
B(\nu_e,T_d)}
\label{eq:mdust}
\end{equation}
where $\nu_e$ and $\nu_{obs}$ are the frequencies at which the
radiation is emitted and observed, $T_d$ is the dust temperature,
$D_L$ is the luminosity distance of the source, computed for a flat
cosmology with a cosmological constant according to \citet{Pen:1999},
$S(\nu_{obs})$ is the observed flux and $B(\nu_e,T_d)$ is the Planck
function.
\begin{table*}
\caption{Optical magnitude and intrinsic luminosity for of subset of
the CCSN sample. The ugriz magnitudes have been retrieved from the
SDSS/DR6 database. The rest-frame B luminosities (L$_{\rm {rest}}^B$) have been computed
relying on the known redshifts.}  
\centering
\begin{tabular}{lllllll}
\hline\hline
Name & u & g & r & i & z & L$_{\rm {rest}}^B$ \\
(1) & (2) & (3)& (4) & (5) & (6) & (7)\\
units & mag & mag & mag & mag & mag & L$_*^B$\\\hline
sn1999fl &22.89$\pm$ 0.45 &23.43$\pm$  0.29   &22.04$\pm$ 0.14  &20.16$\pm$ 0.04  &18.92$\pm$ 0.05& 0.7\\
sn2001ek &23.69$\pm$ 1.79 &23.33$\pm$  0.48   &21.59$\pm$ 0.16  &20.70$\pm$ 0.11 &19.93$\pm$  0.19& 0.6-2.5\\
sn1997ev &21.91$\pm$ 0.35 &21.38$\pm$  0.08   &20.38$\pm$ 0.05  &19.93$\pm$ 0.06 &19.42$\pm$  0.14& 9.2\\
sn1997ew &22.48$\pm$ 0.49 &20.31$\pm$  0.03   &18.93$\pm$ 0.01  &18.43$\pm$ 0.01  &18.06$\pm$ 0.03& 87\\
sn2001ct &23.41$\pm$ 1.08 &23.91$\pm$  0.71   &22.11$\pm$ 0.24  &21.89$\pm$ 0.37  &22.34$\pm$ 1.27& 2.0\\
sn2001gl &22.65$\pm$ 0.36 &19.98$\pm$  0.02   &18.61$\pm$ 0.01  &17.94$\pm$ 0.01  &17.59$\pm$ 0.01& 28\\
\hline
\end{tabular}
\\Note: we consider L$_*^B=2.3\times 10^9 L_\odot = 1.9 \times 10^{10} L^B_\odot$.
\label{tab:sdss}
\end{table*}
\begin{figure}[ht]
\centering
\includegraphics[width=0.5\textwidth]{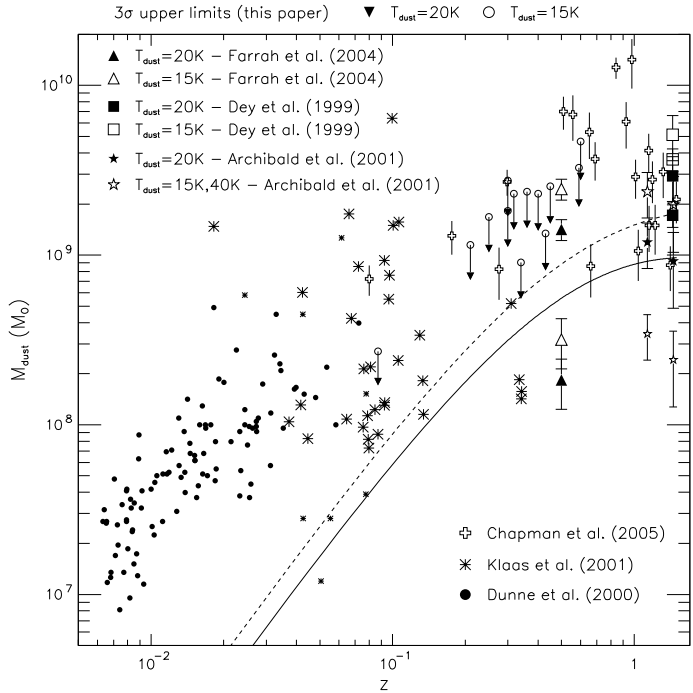}
\caption{Dust mass-redshift diagram. Upper limits obtained for the
dust mass of CCSN hosts are superimposed on dust mass estimates
derived for other galaxies in this redshift range. We derive from our
250GHz measurements (provided in Table \protect\ref{tab:sources})
upper values (3$\sigma$) on the corresponding dust masses for
$T_{dust}=20$K (filled inverted triangle symbols) and $T_{dust}=15$K
(open circle symbols). Lines connect these 2 upper values. The other
points correspond to dust masses derived from 350-GHz (detected)
fluxes (assuming $T_{dust}=20$K and $T_{dust}=15$K when indicated
so). The upright triangle symbols correspond to the SN-Ia host
detected by Farrah et al. (2004)\protect\nocite{Farrah:2004}, and the
mean flux derived from the rest of their SN-Ia host sample at
\protect$z=0.5$. The full (resp. dashed) line correspond to a 1\,mJy
($1\sigma$) sensitivity at 1.2mm for $T_{d}\sim 20$K (resp. $T_{d}\sim
15$K). See text for more details.}
\label{fig:snmm}
\end{figure}

Relying on Eq. \ref{eq:mdust} and the previous parameters, we have
converted our 250\,GHz continuum measurements into dust mass upper
limits and superimposed them on various (350\,GHz-based) dust mass
estimates derived in this redshift range in Fig. \ref{fig:snmm}. Note
that as shown by Seaquist et al. (2004)\protect\nocite{Seaquist:2004}
for the SLUGS data \protect\citep{Dunne:2000} the estimates based on
350-GHz data are overestimated by 25-38$\%$ due to the contamination
by the CO(3-2) line, {which might be counter-balanced by an
increased sensitivity to hot dust components.} We have displayed
the various dust mass estimates obtained at $z<1.5$. The full bullets
correspond to dust mass $M_d^{cold}$ obtained for IRAS Bright Galaxies
($L_{IR}<10^{12} L_\odot$) with a two-component-temperature fit to the
data, assuming a cold $T_d=20K$ and $\beta=2$
\protect\citep{Dunne:2000}. The asterisk symbols provide the dust
masses based on multiple modified blackbody with $\beta=2$ for bright
ultraluminous infrared galaxies
\protect\citep{Klaas:2001}. The largest (resp. smallest) asterisk
symbols correspond to galaxies with $L_{IR}>10^{12} L_\odot$
(resp. $L_{IR}\le 10^{12} L_\odot$). The crosses correspond to
submillimetre galaxies from \citet{Chapman:2005}. For the other data
sets, we derive the dust mass with the 350\,GHz flux measurements. The
star symbols correspond to radio-galaxies \citep{Archibald:2001} for
which the synchrotron emission has been removed from the 350-GHz
fluxes. (Note that the small-size star symbols refer to
$T_{dust}=40$K, favoured by the authors.) The square symbols
correspond to the galaxy HR10 \citep{Dey:1999}. The upright triangle
symbols indicate the SN-Ia hosts detections by Farrah et
al. (2004)\nocite{Farrah:2004}.

Figure \ref{fig:snmm} shows that our upper limits are {consistent}
with the other data detected at $z<1.5$. We can conclude that among
our sample of CCSN hosts, none contains a very large amount of dust
($M_{dust} > 10^9 M_\odot$). {The error-weighted mean flux
corresponds to a dust mass of $1.4\pm 2.2 \times 10^8 M_\odot$, which is
comparable with the \citet{Farrah:2004} sample and the upper
distribution of the local SLUGS sources of \citet{Dunne:2000}.}

\subsection{Optical fluxes}
A sub-set (6/13) of our galaxies has been observed by the SDSS. We
retrieve the corresponding ugriz photometry and derived the intrinsic
B luminosity as displayed in Table \ref{tab:sdss}. We can see the
diversity of the host galaxies on our sample: very massive systems
(87\,L$_*^B$) as well as normal galaxies (0.7\,L$_*^B$) are present,
with an average B luminosity of 21\,L$_*^B$. This simply reflects the
fact that these CCSN are detected in all types of galaxies containing
gas and obeying the selection criteria.
 
\section{Conclusion}
\label{sec:summ}
We have observed 13 high-z CCSN host galaxies at 250\,GHz and detected
no signal. {We can exclude that individual galaxies contain more
than about $10^9 M_\odot$ of dust, while the error-weighted average
dust mass is $1.4\pm 2.2 \times 10^8 M_\odot$ for the whole sample.
Our results are compatible with those of \citet{Farrah:2004} only if
the strong source detected by these authors is excluded, while we find
a good agreement with the results of \citet{Clements:2005}. The B
luminosities of a subset of our sample present a significant diversity
of the hosts. This study shows that CCSN host galaxies are typical of
galaxies observed in the local Universe \citep[e.g.][]{Kauffmann:2003}
and do not belong to the brightest SMG population.}

\begin{acknowledgements} 
We are very grateful to the anonymous referee for his valuable and
detailled comments. We thank Carl Pennypacker who motivated this work
during his sabbatical leave at LERMA in 2003. ALM kindly acknowledges
Axel Weiss for his help during the Pool observations at IRAM, and
Robert Zylka for his efficient introduction to the MOPSI software. ALM
thanks J.-F. Lestrade for helpful discussions. 

Photometry of 6 galaxies has been retrieved from SDSS/Data Release 6.
Funding for the SDSS and SDSS-II has been provided by the Alfred
P. Sloan Foundation, the Participating Institutions, the National
Science Foundation, the U.S. Department of Energy, the National
Aeronautics and Space Administration, the Japanese Monbukagakusho, the
Max Planck Society, and the Higher Education Funding Council for
England. The SDSS Web Site is http://www.sdss.org/.
The SDSS is managed by the Astrophysical Research Consortium for the
Participating Institutions. The Participating Institutions are the
American Museum of Natural History, Astrophysical Institute Potsdam,
University of Basel, University of Cambridge, Case Western Reserve
University, University of Chicago, Drexel University, Fermilab, the
Institute for Advanced Study, the Japan Participation Group, Johns
Hopkins University, the Joint Institute for Nuclear Astrophysics, the
Kavli Institute for Particle Astrophysics and Cosmology, the Korean
Scientist Group, the Chinese Academy of Sciences (LAMOST), Los Alamos
National Laboratory, the Max-Planck-Institute for Astronomy (MPIA),
the Max-Planck-Institute for Astrophysics (MPA), New Mexico State
University, Ohio State University, University of Pittsburgh,
University of Portsmouth, Princeton University, the United States
Naval Observatory, and the University of Washington.
\end{acknowledgements}
%\begin{thebibliography}{} \end{thebibliography}
%\bibliography{bibli}

\end{document}